\documentclass[12pt]{amsart}
\usepackage{amssymb}
\usepackage{txfonts}
\usepackage{amsfonts}
\usepackage{mathrsfs}
 \theoremstyle{definition}
 
 \theoremstyle{remark}

 \numberwithin{equation}{section}

\begin{document}
\title[The RCCN criterion]
{The RCCN criterion of separability for states in
infinite-dimensional quantum systems}

\author{Yu Guo}
\address{Department of Mathematics, Shanxi University, Taiyuan, 030006, China;
Department of Mathematics, Shanxi Datong University, Datong, 037009, China.}
 \email{guoyu3@yahoo.com.cn}

\author{Jinchuan Hou}
\address{Department of Mathematics, Taiyuan University of Technology, Taiyuan 030024, P. R. China;
Department of Mathematics, Shanxi University, Taiyuan, 030006, P. R.
China}
 \email{jinchuanhou@yahoo.com.cn}


\begin{abstract}

In this paper,  the realignment criterion and the RCCN criterion of
separability for states in infinite-dimensional bipartite quantum
systems are established.  Let $H_A$ and $H_B$ be complex Hilbert
spaces with $\dim H_A\otimes H_B=+\infty$. Let $\rho$ be a state on $H_A\otimes H_B$ and $\{\delta_k\}$
be the Schmidt coefficients of $\rho$  as a vector in the Hilbert
space ${\mathcal C}_2(H_A)\otimes{\mathcal C}_2(H_B)$. We introduce
the realignment operation $\rho^R$ and the computable cross norm
$\|\rho\|_{\rm CCN}$ of $\rho$ and show that, if $\rho$ is
separable, then $\|\rho^{R}\|_{\rm Tr}=\|\rho\|_{\rm
CCN}=\sum\limits_k\delta_k\leq1.$ In particular, if $\rho$ is a pure
state, then $\rho$ is separable if and only if $\|\rho^{R}\|_{\rm
Tr}=\|\rho\|_{\rm CCN}=\sum\limits_k\delta_k=1$.

\end{abstract}

\maketitle

\section{Introduction}

The quantum entanglement is one of the most striking features of the
quantum mechanics and it is used as a physical resource for
communication information processing \cite{MN}. Consequently, the
detection of entanglement, that is, distinguishing  separable and
entangled states, has been investigated extensively
\cite{CK,GY,GM,HM1,HP1,HJ,MA,PA,RO1,RO2,RO3,ZZ}. However, in spite
of the considerable effort, no necessary-sufficient criterion that
is practically implementable is known so far even though in
finite-dimensional bipartite quantum systems. The case of
infinite-dimensional systems can't be neglected since they do exist
in the quantum world \cite{Su,Len}. Therefore, how to recognize the
separability of states in infinite-dimensional systems is a more
difficult problem that is of both fundamental and practical
importance within quantum mechanics and quantum information theory.

It is known that, a density operator $\rho$ (i.e., a positive
trace-one operator) acting on a separable Hilbert space $H=H_A\otimes H_B$
describing the state of two quantum systems $\rm A$ and $\rm{B}$, is
called \emph{separable} if it can be written as a convex combination
$$\rho=\sum\limits_ip_i\rho_i^A\otimes\rho_i^B,\ \ \sum\limits_ip_i=1, \
p_i\geq0 \eqno{(1)}$$ or can be approximated in the trace norm by
the states of the above form \cite{HP2,Wer}, where $\rho_i^A$ and
$\rho_i^B$ are (pure) states in the subsystems ${\rm A}$ and ${\rm
B}$ which are described by the complex Hilbert spaces $H_A$ and
$H_B$, respectively. Otherwise, $\rho$ is called \emph{entangled}.
Let $\mathcal{S}_{s-p}$ be the set of all separable pure states. It
is shown in \cite{Hol} that, any separable state $\rho$ admits a
representation of the Bochner integral
$$\rho=\int_{\mathcal{S}_{s-p}}\varphi(\rho^A\otimes\rho^B)d\mu(\rho^A\otimes\rho^B),\eqno{(2)}$$
where $\mu$ is a Borel probability measure on $\mathcal{S}_{s-p}$, $\rho^A\otimes\rho^B\in\mathcal{S}_{s-p}$
and  $\varphi:\mathcal{S}_{s-p}\rightarrow\mathcal{S}_{s-p}$ is a
measurable function. Particularly, if $\dim H_A\otimes H_B<+\infty$,
then a state $\rho$ acting on $H_A\otimes H_B$ is
separable if and only if $\rho$ can be written as \cite{HP2}
$$\rho=\sum\limits_{i=1}^np_i\rho_i^A\otimes\rho_i^B,\eqno{(3)}$$
where $\rho_i^A$ and $\rho_i^B$ are pure states in the subsystems
${\rm A}$ and ${\rm B}$, respectively, and where $p_i\geq0$ with
$\sum\limits_{i=1}^np_i=1$ and $n\leq(\dim H_A\otimes H_B)^2$. In the
infinite-dimensional case, there exists separable state that can not
be written in the form $\sum\limits_{i=1}^{+\infty}
p_i\rho_i^A\otimes\rho_i^B$ with $\sum\limits_{i=1}^{+\infty} p_i=1$
\cite{Hol}.

For the finite-dimensional bipartite quantum systems, K. Chen and
L.-A. Wu proposed the realignment criterion   in \cite{CK}, which
reads as: if $\rho$ is a separable state of the bipartite quantum
system, then the trace norm of the realignment matrix of $\rho$ is
not larger than 1. A short later, O.Rudolph proved in
\cite{RO2} that if $\rho$ is a state of the bipartite quantum
system, then the computable cross norm of $\rho$ equals the trace
norm of the realignment matrix of $\rho$.  This result, combining
the result in \cite{CK}, is called the \emph{realignment criterion
or computable cross norm} \emph{criterion} (or RCCN criterion
briefly) \cite{CK,RO1, RO2}.  Then, a natural problem is arisen:
whether or not there is a counterpart result for the
infinite-dimensional bipartite quantum systems? We find that the
answer is `yes'. The aim of the present paper is to establish the realignment criterion and the
RCCN criterion for the infinite-dimensional bipartite quantum
systems.

The paper is organized as follows. In section 2, we summarize the
studies on the realignment criterion and the RCCN criterion for finite-dimensional bipartite
quantum systems, which enlightens the way how to generalize the
conception of realignment to the  infinite-dimensional case. Section
3 devotes to generalizing the notion of the realignment operation to
the infinite-dimensional systems, and presenting the  realignment
criterion and the RCCN criterion for infinite-dimensional bipartite
quantum systems. Let $H_A$ and $H_B$ be Hilbert spaces. We introduce
three equivalent definitions of the realignment operation from the
Hilbert-Schmidt class ${\mathcal C}_2(H_A\otimes H_B)$ into
${\mathcal C}_2(H_B\otimes H_B, H_A\otimes H_A)$ and reveal that the
realignment operation
$$T\mapsto T^R\eqno{(4)}$$
is an isometry with respect to the Hilbert-Schmidt norm $\|\cdot\|_2$.
 Let
$\rho$ be a state on $H_A\otimes H_B$ and $\{\delta_k\}$ be the
Schmidt coefficients of $\rho$   regarded as a vector in the Hilbert space
${\mathcal C}_2(H_A)\otimes{\mathcal C}_2(H_B)$. We show that, if
$\rho$ is separable, then $\|\rho^{R}\|_{\rm Tr}=\|\rho\|_{\rm
CCN}=\sum\limits_k\delta_k\leq1.$ In particular, if $\rho$ is a pure
state, then $\rho$ is separable if and only if $\|\rho^{R}\|_{\rm
Tr}=\|\rho\|_{\rm CCN}=\sum\limits_k\delta_k=1$ (Criteria 3.3 and
3.7). Thus $\|\rho^{R}\|_{\rm Tr}=\|\rho\|_{\rm
CCN}=\sum\limits_k\delta_k>1$ signals the entanglement of $\rho$.
The RCCN criterion just provides a necessary condition for
separability \cite{RO2}. However, the RCCN criterion can detect many
states with positive partial transpose (PPT) \cite{RO1,RO2},
i.e., the so-called PPT states (which are bound entangled states).
Several examples are given to illustrate the relations between the
RCCN criterion and the PPT criterion. They show that the
infinite-dimensional RCCN criterion can also detect some PPT states
as desired (see Examples 3.8-3.10). A final conclusion is included
in the last section.

We fix some notations. Throughout the paper we use the Dirac's
symbols. ${\mathbb R}$, $\mathbb C$ and $\mathbb N$ stand for the set of all real numbers, the set of all complex numbers and the set of all
nonnegative integers, respectively. The bra-ket
notation, $\langle\cdot|\cdot\rangle$ stands for the inner product
in the given Hilbert spaces, i.e., $H_A\otimes H_B$, $H_A$, or
$H_B$. The set of all bounded linear operators on some Hilbert space $H$ is denoted by
$\mathcal{B}(H)$, the set of trace class operators on $H$ is denoted
by $\mathcal{T}(H)$ and the set of all Hilbert-Schmidt class
operators on $H$ is denoted by $\mathcal{C}_2(H)$.
$A\in\mathcal{B}(H)$ is self-adjoint if $A^\dagger=A$ ($A^\dagger$
stands for the adjoint operator of $A$); $A$ is said to be positive,
denoted by $A\geq0$, if $A^\dagger=A$ and
$\langle\psi|A|\psi\rangle\geq0$ for all $|\psi\rangle\in H$. $A^T$
stands for the transpose of the operator $A$, $\|\cdot\|_{\rm
Tr}$ denotes the trace norm and $\|\cdot\|_2$ denotes the
Hilbert-Schmidt norm, i.e., $\|A\|_{\rm Tr}={\rm Tr}((A^\dagger
A)^{\frac{1}{2}})$ and $\|A\|_2=({\rm Tr}(A^\dagger
A))^{\frac{1}{2}}$. By $\mathcal{S}(H_A)$,
$\mathcal{S}(H_B)$ and
$\mathcal{S}(H_A\otimes H_B)$ we denote the sets of
states on $H_A$, $H_B$ and $H_A\otimes H_B$, respectively. By
$\mathcal{S}_{sep}$ we denote the
set of all separable states in $\mathcal{S}(H_A\otimes H_B)$. A state $\rho$ is
called a pure state if ${\rm Tr}(\rho^2)=1$ and is called a mixed
state  if ${\rm Tr}(\rho^2)<1$ as usual. We also call a unit vector
$|\psi\rangle\in H_A\otimes H_B$ a pure state which is corresponding
to the density operator $\rho=|\psi\rangle\langle\psi|$.  We fix in
the `local Hilbert space' $H_A$, $H_B$ orthonormal bases
$\{|m\rangle\}_{m=1}^{N_A}$ and $\{|\mu\rangle\}_{\mu=1}^{N_B}$,
where $N_A=\dim H_A$  and $N_B=\dim H_B$,   respectively (note that
we use Latin indices for the subsystem $\rm{A}$ and the Greek
indices for the subsystem $\rm{B}$. Also, $N_A$ and $N_B$ may be
$+\infty$). Then, a vector $|\psi\rangle\in H_A\otimes H_B$ can be
written as $|\psi\rangle=\sum_{m,\mu}d_{m\mu}|m\rangle|\mu\rangle\in
H_A\otimes H_B$. Let $D_{\psi}=(d_{m\mu})$ (or $[d_{m\mu}]$) be the
coefficient operator of $|\psi\rangle$. Remark that
$D_{\psi}=(d_{m\mu})$ can be regarded as an operator from $H_B$ into
$H_A$ and it is a Hilbert-Schmidt class operator with the
Hilbert-Schmidt norm $\|D_{\psi}\|_2=\||\psi\rangle\|$. We write
$\bar{D}=(\bar{d_{m\mu}})$, where $\bar{d_{m\mu}}$ is the complex
conjugation of $d_{m\mu}$. The partial transpose
of $\rho\in\mathcal{S}(H_A\otimes H_B)$ with respect to the subsystem ${\rm B}$
(resp. ${\rm A}$) is denoted by $\rho^{T_B}$ (resp. $\rho^{T_A}$),
that is, $\rho^{T_B}=(I\otimes {\bf T})\rho$ (resp. $\rho^{T_A}=(
{\bf T}\otimes I)\rho$), where $\bf T$ is the map of taking
transpose with respect to the given orthonormal basis.

\section{The RCCN criterion for finite-dimensional systems}

To find a way of generalizing the notion of the realignment of a
block matrix to that of an operator matrix acting on an infinite-dimensional
Hilbert space, in this section, we summarize some known facts about
the realignment criterion and the related RCCN criterion for finite-dimensional bipartite quantum
systems in references \cite{CK,RO2,AL1,AL2,ZB} and discuss them
briefly. Assume that ${\rm dim}H_A=N_A$ and ${\rm dim}H_B=N_B$ are
finite throughout this section.

Firstly, we recall the definition of the realignment operation for
the $N_AN_B\times N_AN_B$ matrices, i.e., the $N_A\times N_A$ block
matrices with each block is of size $N_B\times N_B$.  Recalling
that, for a $N_A\times N_A$ block matrix $T=(B_{ij})_{N_A\times
N_A}$ with each block $B_{ij}$ of the size $N_B\times N_B$, $1\leq
i$,$j\leq N_A$, the row realignment matrix $T^{R}$ of $T$ is defined
as
$$\begin{array}{rl}T^{R}=&[({\rm vec}(B_{11}))^{T},\dots,({\rm vec}(B_{1N_A}))^{T},
\dots,\\&({\rm vec}(B_{N_A1}))^{T},\dots,({\rm
vec}(B_{N_AN_A}))^{T}]^{T},\end{array}\eqno{(5)}$$ which is a
$N_A^2\times N_B^2$ matrix, where for a given $X=[x_{ij}]$ with
$1\leq i\leq s$ and $1\leq j\leq t$, ${\rm vec}(X)$ is defined by
$${\rm vec}(X)
=[x_{11},\dots,x_{1t},x_{21},\dots,x_{2t},\dots,x_{s1},\dots,x_{st}].$$
For example, in the case of a two-qubit system, let
$$\rho=\left(\begin{array}{cc}B_{11}&B_{12}\\B_{21}&B_{22}\end{array}\right)=\left(\begin{array}{cc|cc}
\rho_{11}&\rho_{12}&\rho_{13}&\rho_{14}\\
\rho_{21}&\rho_{22}&\rho_{23}&\rho_{24}\\ \hline
\rho_{31}&\rho_{32}&\rho_{33}&\rho_{34}\\
\rho_{41}&\rho_{42}&\rho_{43}&\rho_{44}\end{array}\right),$$ where
$B_{ij}$s are operators on the space associated with the second
system. Then the row realignment matrix of $\rho$ (ref. \cite{AL2})
is
$$\rho^R=\left(\begin{array}{cccc}\rho_{11}&\rho_{12}&\rho_{21}&\rho_{22}\\
\rho_{13}&\rho_{14}&\rho_{23}&\rho_{24}\\
\rho_{31}&\rho_{32}&\rho_{41}&\rho_{42}\\
\rho_{33}&\rho_{34}&\rho_{43}&\rho_{44}\end{array}\right).$$ It is
clear that the realignment operation $T\mapsto T^R$ is a linear map,
that is, $(\alpha T+\beta S)^R=\alpha T^R+\beta S^R$, $\alpha$, $\beta\in\mathbb{C}$.

The so-called realignment criterion due to Chen and Wu \cite{CK} is
the following

{\bf The realignment criterion for finite-dimensional bipartite
systems.} {\it Assume that $H_A$ and $H_B$ are of finite-dimensions
and $\rho\in{\mathcal S}(H_A\otimes H_B)$ is a state. If $\rho$ is
separable, then $\|\rho^R\|_{\rm Tr}\leq 1$.}

The realignment criterion presents a quite strong necessary condition for
separability which is easily performed and independent to the
well-known PPT criterion. However, the above definition of the
realignment operation cannot be generalized to the
infinite-dimensional cases. Fortunately, there are several different
definitions of the realignment operation that are equivalent to
each other. This allows us to find ways of generalizing the
realignment operation to infinite-dimensional cases.

With respect to a fixed product basis $\{|m\rangle|\mu\rangle\}$ of
$H_A\otimes H_B=\mathbb{C}^{N_A}\otimes \mathbb{C}^{N_B}$, every
operator $A\in{\mathcal B}(H_A\otimes H_B)$  can be written in the
form $A =[a_{m\mu,n\nu}]$, where the entry $a_{m\mu,n\nu}=\langle
m|\langle\mu|A|n\rangle|\nu\rangle$, the double indices
$(m\mu)\leftrightarrow(m-1)N_B+\mu$ and
$(n\nu)\leftrightarrow(n-1)N_B+\nu$ refer respectively to rows and
columns of matrix $A$. Then we have \cite{AL2}
$$A^R=[\tilde{a}_{mn,\mu\nu}],\
\tilde{a}_{mn,\mu\nu}=a_{m\mu,n\nu},\eqno{(6)}$$ where the double
indices $(mn)\leftrightarrow(m-1)N_A+n$ and
$(\mu\nu)\leftrightarrow(\mu-1)N_B+\nu$ refer respectively to rows
and columns of matrix $A^R$. For the above example $\rho$ in the
case of a two-qubit system, using the double indices, we may write
$$\rho=\left(\begin{array}{cccc}
\rho_{11,11}&\rho_{11,12}&\rho_{11,21}&\rho_{11,22}\\
\rho_{12,11}&\rho_{12,12}&\rho_{12,21}&\rho_{12,22}\\
\rho_{21,11}&\rho_{21,12}&\rho_{21,21}&\rho_{21,22}\\
\rho_{22,11}&\rho_{22,12}&\rho_{22,21}&\rho_{22,22}\end{array}\right)$$
and then
$$\rho^R=\left(\begin{array}{cccc}
\tilde{\rho}_{11,11}&\tilde{\rho}_{11,12}&\tilde{\rho}_{11,21}&\tilde{\rho}_{11,22}\\
\tilde{\rho}_{12,11}&\tilde{\rho}_{12,12}&\tilde{\rho}_{12,21}&\tilde{\rho}_{12,22}\\
\tilde{\rho}_{21,11}&\tilde{\rho}_{21,12}&\tilde{\rho}_{21,21}&\tilde{\rho}_{21,22}\\
\tilde{\rho}_{22,11}&\tilde{\rho}_{22,12}&\tilde{\rho}_{22,21}&\tilde{\rho}_{22,22}
\end{array}\right)
=\left(\begin{array}{cccc}
\rho_{11,11}&\rho_{11,12}&\rho_{12,11}&\rho_{12,12}\\
\rho_{11,21}&\rho_{11,22}&\rho_{12,21}&\rho_{12,22}\\
\rho_{21,11}&\rho_{21,12}&\rho_{22,11}&\rho_{22,12}\\
\rho_{21,21}&\rho_{21,22}&\rho_{22,21}&\rho_{22,22}\end{array}\right).$$

The operation of realignment can also be defined in another
alternative way \cite{RO2}. For a $N_A\times N_A$ matrix
$A=[a_{mn}]\in\mathcal{B}(H_A)$ (resp. $N_B\times N_B$ matrix
$B=[b_{\mu\nu}]\in\mathcal{B}(H_B)$) in terms of the basis
$\{|m\rangle\}$ (resp. $\{|\mu\rangle\}$ ), regard $A$ (resp. $B$)
as a vector $|A\rangle=\sum\limits_{m,n}a_{mn}|m\rangle|n\rangle$ in
${\mathbb C}^{N_A^2}$ (resp.
$|B\rangle=\sum\limits_{\mu,\nu}b_{\mu\nu}|\mu\rangle|\nu\rangle$ in
${\mathbb C}^{N_B^2}$). If $\rho=\sum\limits_{k=1}^sA_k\otimes B_k\in{\mathcal S}(H_A\otimes H_B)$,
then
$$\rho^R=\sum\limits_{k=1}^s|A_k\rangle\langle B_k|,\eqno{(7)}$$
{\it where $\langle B_k|$ denotes the transpose of} $|B_k\rangle$ (not the conjugate transpose as usual),
$k=1$, 2, $\dots$, $s$. In particular, for any pure state
$\rho_\psi=|\psi\rangle\langle\psi|$, write
$|\psi\rangle=\sum_{m,\mu}d_{m\mu}|m\rangle|\mu\rangle$ and
$D=[d_{m\mu}]$, then by \cite{AL2},
$$\rho_\psi^R=D\otimes\bar{D}.\eqno{(8)}$$ It follows that, for any mixed
state   $\rho=\sum\limits_{i=1}^tp_i\rho_i$, where $p_i\geq0$,
$\sum\limits_{i=1}^tp_i=1$ and $\rho_i$ are pure states of the
bipartite system, $i=1$, 2, $\dots$, $t$, $\rho^R$
can be defined to be
$$\rho^R=\sum\limits_{i=1}^tp_i\rho_i^R
=\sum\limits_{i=1}^tp_iD_i\otimes\bar{D_i},\eqno{(9)}$$ where
$\rho_i=|\psi_i\rangle\langle\psi_i|$ with $|\psi_i\rangle=\sum
d^{(i)}_{m\mu}|m\rangle|\mu\rangle$ and $D_i=[d^{(i)}_{m\mu}]$ is
the coefficient matrix of $|\psi_i\rangle$ \cite{AL2}.

Similarly, the column realignment matrix of $\rho$, denoted by
$\rho^{R^c}$, was defined in  \cite{CK}. For the two-qubit state $\rho$ mentioned above, we have
$$\rho^{R^c}
=\left(\begin{array}{cccc}\rho_{11}&\rho_{21}&\rho_{12}&\rho_{22}\\
\rho_{31}&\rho_{41}&\rho_{32}&\rho_{42}\\
\rho_{13}&\rho_{23}&\rho_{14}&\rho_{24}\\
\rho_{33}&\rho_{43}&\rho_{34}&\rho_{44}\end{array}\right).$$

It is easy to check that,
$$\rho^{R^c}
=[\tilde{\rho}_{\nu\mu,nm}]^{T},\ \tilde{\rho}_{\nu\mu,nm}
=\rho_{m\mu,n\nu};\eqno{(10)}$$ if $\rho=\sum\limits_kA_k\otimes
B_k$ with $A_k=\sum\limits_{m,n}a_{mn}^{(k)}|m\rangle\langle n|$ and
$B_k=\sum\limits_{\mu,\nu}b_{\mu\nu}^{(k)}|\mu\rangle\langle\nu|$,
then
$$\rho^{R^c}
=\sum\limits_{k=1}^s|\tilde{A}_k\rangle\langle\tilde{B}_k|,\eqno{(11)}$$
where
$|\tilde{A}_k\rangle=\sum\limits_{m,n}a_{mn}^{(k)}|n\rangle|m\rangle$
and $|\tilde{B}_k\rangle
=\sum\limits_{\mu,\nu}b_{\mu\nu}^{(k)}|\nu\rangle|\mu\rangle$,
$k=1$, $2$, $\dots$, $s$; if $\rho=\sum\limits_{i=1}^t p_i\rho_i$,
then
$$\rho^{R^c}=\sum\limits_{i=1}^tp_i\rho_i^{R^\prime}
=\sum\limits_{i=1}^tp_iD_i^T\otimes\bar{D_i}^T,\eqno{(12)}$$ where
$\rho_i$, $p_i$ and $D_i$ defined as in Eq.(9). For instance, using
the double indices, the column realignment matrix of the
example $\rho$ mentioned above is
$$\rho^{R^c}=\left(\begin{array}{cccc}
\tilde{\rho}_{11,11}&\tilde{\rho}_{11,12}&\tilde{\rho}_{11,21}&\tilde{\rho}_{11,22}\\
\tilde{\rho}_{12,11}&\tilde{\rho}_{12,12}&\tilde{\rho}_{12,21}&\tilde{\rho}_{12,22}\\
\tilde{\rho}_{21,11}&\tilde{\rho}_{21,12}&\tilde{\rho}_{21,21}&\tilde{\rho}_{21,22}\\
\tilde{\rho}_{22,11}&\tilde{\rho}_{22,12}&\tilde{\rho}_{22,21}&\tilde{\rho}_{22,22}
\end{array}\right)^T
=\left(\begin{array}{cccc}
\rho_{11,11}&\rho_{12,11}&\rho_{11,12}&\rho_{12,12}\\
\rho_{21,11}&\rho_{22,11}&\rho_{21,12}&\rho_{22,12}\\
\rho_{11,21}&\rho_{12,21}&\rho_{11,22}&\rho_{12,22}\\
\rho_{21,21}&\rho_{22,21}&\rho_{21,22}&\rho_{22,22}\end{array}\right).$$

The singular values of $\rho^R$ and $\rho^{R^c}$ are equal
\cite{ZB}. In fact, let
$F_A=\sum\limits_{m,n=1}^{N_A}|m\rangle\langle
n|\otimes|n\rangle\langle m|$ and
$F_B=\sum\limits_{\mu,\nu=1}^{N_B}|\mu\rangle\langle
\nu|\otimes|\nu\rangle\langle\mu|$; then $F_{A}$ (resp. $F_B$) is a
unitary matrix of size $N_A^2\times N_A^2$ (resp. $N_B^2\times
N_B^2$). $F_A$ and $F_B$ are the so-called swap operators or the
flip operators \cite{TG}. It is easily checked that
$F_A|A_k\rangle=|\tilde{A}_k\rangle$ and
$F_B|B_k\rangle=|\tilde{B}_k\rangle$, $k=1$, 2, $\dots$. It turns
out that
$$\rho^R=F_A\rho^{R^c}F_B.\eqno{(13)}$$
Therefore, we need to consider the row realignment only.

In the following, the realignment of a matrix always refers to the
row realignment of the matrix unless specified.

Note that, for any state $\rho\in\mathcal{S}(H_A\otimes H_B)$, one has
$$\|\rho^R\|_2=\|\rho\|_2\leq\|\rho\|_{\rm Tr}=1.\eqno{(14)}$$

For any $C\in\mathcal{B}(H_A\otimes H_B)$, the \emph{computable
cross norm} of $C$, $\|C\|_{\rm CCN}$, is defined by:
$$\begin{array}{rl}\|C\|_{\rm CCN}:=&\inf\{\sum\limits_{k=1}^s
\|A_k\|_2\|B_k\|_2:C=\sum\limits_{k=1}^sA_k\otimes B_k,\\&
A_k\in\mathcal{B}(H_A), B_k\in\mathcal{B}(H_B)\},\end{array}\eqno{(15)}$$
where the infimum is taken over all finite decompositions of $C$
into a finite sum of simple tensors  \cite{RO2}.

 Notice that the linear space $\mathcal{B}(H_A\otimes H_B)$ can be
considered as a Hilbert space if it is equipped with the (complex)
Hilbert-Schmidt scalar product: $$\langle A|B\rangle:={\rm
Tr}(A^\dagger B), A,B\in\mathcal{B}(H_A\otimes H_B),$$ the
Hilbert-Schmidt norm, $\|\cdot\|_2$, reads as $$\|A\|_2:=({\rm
Tr}(A^\dagger A))^{1/2}.$$ Then, every $\rho\in\mathcal{S}(H_A\otimes H_B)$
can be regarded as a `vector' in the Hilbert space
$\mathcal{B}(H_A\otimes H_B)$ equipped with the Hilbert-Schmidt
inner product. It follows that there
is a Schmidt decomposition of $\rho$:
$$\rho=\sum\limits_{k=1}^r\delta_kE_k\otimes F_k, $$
where the coefficients $\{\delta_k\}$ are positive, $\{E_k\}$,
$\{F_k\}$ are orthonormal sets of Hilbert spaces $\mathcal{B}(H_A)$,
$\mathcal{B}(H_B)$, respectively, and  $r$ is the Schmidt number of
$\rho$. The set of the positive numbers $\{\delta_k\}$ is uniquely
determined by the corresponding vector $\rho$, and they are called
the Schmidt coefficients of $\rho$ \cite{ZB}.

It is showed in \cite{RO2,AL2,ZB} that $$\|\rho\|_{\rm CCN}=\|\rho^{
R}\|_{\rm Tr}=\sum\limits_{k=1}^r\delta_k,\eqno{(16)}$$ where
$\delta_k$ is the Schmidt coefficients of $\rho$.  From this point
of view, this cross norm $\|\rho\|_{\rm CCN}$ of $\rho$  is called
computable cross norm of $\rho$, and the following criterion is
called  \emph{the realignment or computable cross norm criterion}
(the RCCN criterion for short) due to
 \cite{RO2,AL1, AL2}.

{\bf The RCCN criterion for finite-dimensional bipartite quantum
systems}  \ {\it Let $H_A$ and $H_B$ be finite-dimensional Hilbert
spaces and  $\rho\in\mathcal{S}(H_A\otimes H_B)$ be a state. Let $\rho=\sum\limits_{k=1}^r\delta_kE_k\otimes F_k$ be the Schmidt
decomposition of $\rho$  as a vector of ${\mathcal C}_2(H_A)\otimes {\mathcal
C}_2(H_B)$. If $\rho$ is separable, then
$$\|\rho\|_{\rm CCN}=\|\rho^{ R}\|_{\rm
Tr}=\sum\limits_{k=1}^r\delta_k\leq1.\eqno{(17)}$$ In particular, if
$\rho$ is a pure state, then $\rho$ is separable if and only if
$$\|\rho\|_{\rm CCN}=\|\rho^{R}\|_{\rm
Tr}=\sum\limits_{k=1}^r\delta_k=1.\eqno{(18)}$$}

It is known that the RCCN criterion is neither weaker nor stronger
than the PPT criterion \cite{RO2}. Namely, there exist PPT entangled
states which can be detected by the RCCN criterion, while there are
non-PPT entangled states  which can not be detected by the RCCN
criterion (for instance,  certain $d\times d$ Werner states, ref.
\cite{RO1,RO2, HP2}).

\section{The RCCN criterion for  infinite-dimensional systems}

In this section, we will establish the realignment criterion and the RCCN criterion for
infinite-dimensional bipartite quantum systems. Unless specifically
stated, we assume that at least one of $H_A$ and $H_B$ is of
infinite dimension throughout this section.

In \cite{GY}, we proposed a so-called realignment operation for a
given pure state in infinite-dimensional bipartite quantum systems.
For a given fixed product basis $\{|m\rangle|\mu\rangle\}$ of
$H_A\otimes H_B$, every unit vector $|\psi\rangle$ can be written in
$|\psi\rangle=\sum\limits_{m,\mu}d_{m\mu}|m\rangle|\mu\rangle$.
Write $D=(d_{m\mu})$ and $\bar{D}=(\bar{d_{m\mu}})$. Then the
realignment operator of the pure state
$\rho_{\psi}=|\psi\rangle\langle\psi|$ is defined to be
$$\rho_\psi^{R^\prime}=D\otimes\bar{D}.\eqno{(19)}$$
It is straightforward that $\|\rho_\psi^{R^\prime}\|_2=\|D\otimes
\bar{D}\|_2=\|D\|_2\cdot\|\bar{D}\|_2=1$,
$\rho_\psi^{R^\prime}\in\mathcal{C}_2(H_B\otimes H_B,H_A\otimes
H_A)$. As the realignment operation must be linear, we can define a
realignment operation for a mixed state $\rho=\sum_{i=1}^k
p_i\rho_i$ by $\rho^{R^\prime}=\sum_{i=1}^kp_i\rho_i^{R^\prime}$,
where $\rho_i$s are pure states, $k\in{\mathbb N}$ or $k=+\infty$.
This definition obviously coincides with that for finite-dimensional
systems.

 Like the case of finite-dimensions, for
an arbitrarily fixed product  basis $\{|m\rangle|\mu\rangle\}$,
$\rho_\psi$ can be written in an infinite matrix of double indices
$$\rho_\psi=(\rho_{m\mu,n\nu}^\psi),\ \rho_{m\mu,n\nu}^\psi:=\langle
m|\langle\mu|\rho_\psi|n\rangle|\nu\rangle  \eqno{(20)}$$  and we have
$$\rho_\psi^{R^\prime}=D\otimes \bar{D}=(\tilde{\rho}_{mn,\mu\nu}^\psi), \
\tilde{\rho}_{mn,\mu\nu}^\psi=\langle m|\langle
n|\rho^{R^\prime}_\psi|\mu\rangle|\nu\rangle.\eqno{(21)}$$  It is easy to check that
$$\tilde{\rho}^\psi_{mn,\mu\nu}=\rho_{m\mu,n\nu}^\psi.\eqno{(22)}$$

Inspired by Eq.(22), we now give a definition of the realignment
operation. As usual, we denote by ${\mathcal C}_2(H,K)$ (${\mathcal
C}_2(H)$ if $H=K$) the set of all Hilbert-Schmidt operator from the
Hilbert space $H$ into the Hilbert space $K$. That is,
$\mathcal{C}_2(H,K)$ is a Hilbert space with respect to the complex
scalar product $\langle A|B\rangle:={\rm Tr}(A^\dagger B),
A,B\in\mathcal{C}_2(H,K).$ The Hilbert-Schmidt norm of $A$ is
$\|A\|_2=({\rm Tr}(A^\dagger A))^{1/2}$.

{\bf Definition 3.1.}\ {\it Let $T\in\mathcal{C}_2(H_A\otimes H_B)$
and $Z\in\mathcal{C}_2(H_B\otimes H_B,H_A\otimes H_A)$. Let
$\{|m\rangle\}$ and $\{|\mu\rangle\}$ be arbitrarily given
orthonormal bases  of $H_A$ and $ H_B$, respectively. Then  $T$ and
$Z$ can be written respectively in
$$T=(t_{m\mu,n\nu}),\ t_{m\mu,n\nu}=\langle
m|\langle\mu|T|n\rangle|\nu\rangle $$ and $$Z=(z_{mn,\mu\nu}),\
z_{mn,\mu\nu}=\langle m|\langle n|Z|\mu\rangle|\nu\rangle.$$ If
$$z_{mn,\mu\nu}=t_{m\mu,n\nu},\eqno{(23)}$$ we say that $Z$ is
 the realignment operator of $T$, denoted by
$T^R=Z$, with respect to the given bases.}

The realignment operation ${\mathcal R}:\mathcal{C}_2(H_A\otimes
H_B)\rightarrow\mathcal{C}_2(H_B\otimes H_B,H_A\otimes H_A)$ defined
by ${\mathcal R}T=T^R$ as in the Definition 3.1 is an isometry,
namely, ${\mathcal R}$ is linear and $\|{\mathcal
R}T\|_2=\|T^R\|_2=\|T\|_2$ for every $T$. Particularly, for any
$\rho\in\mathcal{S}(H_A\otimes H_B)$, we have
$$\|\rho^R\|_2=\|\rho\|_2\leq\|\rho\|_{\rm Tr}=1.\eqno{(24)}$$

By Eqs.(20)-(22) and Definition 3.1, it follows that, for any mixed
state $\rho=\sum\limits_ip_i|\psi_i\rangle\langle\psi_i|$ where
$\sum\limits_ip_i=1$, $p_i\geq0$,
$\rho_i=|\psi_i\rangle\langle\psi_i|$ are pure states, we have
$$\rho^R=\sum\limits_i p_iD_i\otimes\bar{D}_i=\rho^{R^\prime},\eqno{(25)}$$ where
the series converges in Hilbert-Schmidt norm on
$\mathcal{C}_2(H_B\otimes H_B,H_A\otimes H_A)$, and
$D_i=(d^{(i)}_{m\mu})$ whenever
$|\psi_i\rangle=\sum\limits_{m,\mu}d^{(i)}_{m,\mu}|m\rangle|\mu\rangle$.
Thus the realignment operation defined in Definition 3.1  coincides
with that introduced in \cite{GY}. It is also clear that $\rho^R$ is
independent on the decomposition of $\rho$, that is, if
$\rho=\sum\limits_jq_j|\phi_j\rangle\langle\phi_j|$ is another
decomposition of $\rho$ into an infinite convex combination, then
$\sum\limits_ip_iD_i\otimes\bar{D}_i=\sum\limits_jq_jD_j^\prime\otimes\bar{D}_j^\prime$,
where $D_j^\prime=(d^{(j)}_{m\mu})$ whenever
$|\phi_j\rangle=\sum\limits_{m,\mu}d^{(j)}_{m,\mu}|m\rangle|\mu\rangle$.

{\bf Remark 3.2.}\ (1) It is clear that the realignment operation in
Definition 3.1 is an infinite-dimensional generalization of the row
realignment of a matrix for finite-dimensional case as discussed in
section 2.

(2) The trace norm and the Hilbert-Schmidt norm of the realignment
operator of a state is independent on  the choice of the bases of $H_A$
and $H_B$.

(3) Similarly, we can define the column realignment operation
$T\mapsto T^{R^c}$ by Eq.(10). Then, if
$\rho=\sum_ip_i|\psi_i\rangle\langle\psi_i|$, we have
$\rho^{R^c}=\sum_ip_iD_i^T\otimes\bar{D_i}^T$, where $p_i$, $D_i$
are the same as the ones mentioned above. Let
$F_A:=\sum_{m,n}|m\rangle|n\rangle\langle n|\langle m|$ and
$F_B:=\sum_{\mu,\nu}|\mu\rangle|\nu\rangle\langle\nu|\langle\mu|$.
It turns out that $F_{A/B}^\dagger=F_{A/B}$,
$F_{A/B}F_{A/B}^\dagger=I_{A/B}$ and $\rho^{R^c}=F_A\rho^RF_B$,
where $I_{A/B}$ is the identity operator on $H_{A/B}$. It follows
that $\|\rho^R\|_{\rm Tr}=\|\rho^{R^c}\|_{\rm Tr}$. Thus, it is
sufficient to discuss the row realignment operation only.

In what follows, we will show that the
set of the realignment operators of the separable states are trace class
operators from $H_B\otimes H_B$ into $H_A\otimes H_A$. In fact, we
have

{\bf Criterion 3.3.}\ ({\bf The realignment criterion for
infinite-dimensional bipartite quantum systems})\ {\it If
$\rho\in\mathcal{S}(H_A\otimes H_B)$ is separable, then
$$\|\rho^R\|_{\rm Tr}\leq1.\eqno{(26)}$$ In particular, if $\rho$ is
a pure state, then $\rho$ is separable if and only if
$$\|\rho^R\|_{\rm Tr}=1.\eqno{(27)}$$}

{\bf Proof.} \ The last assertion was already proved in \cite{GY},
that is,  for the case that $\rho$ is a pure state, $\rho$ is
separable if and only if $\|\rho^R\|_{\rm Tr}=1$.

If $\rho$ is a separable mixed state, then by Eq.(2), there exist a
Borel probability measure $\mu$ on $\mathcal{S}_{s-p}$ and a
measurable function
$\varphi:\mathcal{S}_{s-p}\rightarrow\mathcal{S}_{s-p}$ such that
$\rho$ has a representation of the Bochner integral
$$\rho=\int_{\mathcal{S}_{s-p}}\varphi(\rho^A\otimes\rho^B)d\mu(\rho^A\otimes\rho^B),\quad\rho^A\otimes\rho^B\in\mathcal{S}_{s-p} .\eqno{(28)}$$
It is known that, from the definition of the Bochner integral, there
exists a sequence of step functions $\{\varphi_n\}$ such that
$$\varphi(\rho^A\otimes\rho^B)=\lim\limits_{n\rightarrow\infty}\varphi_n(\rho^A\otimes\rho^B)$$ with respect to the trace norm,
where
$$\varphi_n(\rho^A\otimes\rho^B)=\sum\limits_{i=1}^{k_n}\chi_{E_i}(\rho^A\otimes\rho^B)\rho^A_i\otimes\rho^B_i,$$
$\chi_{E_i}(\cdot)$ is the characteristic function of $E_i$ and
$\{E_i\}_{i=1}^{k_n}$ is a partition of $\mathcal{S}_{s-p}$. Thus
$$\rho=\lim\limits_{n\rightarrow\infty}\sum\limits_{i=1}^{k_n}\mu(E_i)\rho^A_i\otimes\rho^B_i$$
with respect to the trace norm, as well as with respect to the Hilbert-Schmidt norm. Because the realignment operation is continuous,
we have $$\begin{array}{rl}\rho^R&=\lim\limits_{n\rightarrow\infty}\sum\limits_{i=1}^{k_n}\mu(E_i)(\rho^A_i\otimes\rho^B_i)^R\\
&=\lim\limits_{n\rightarrow\infty}\int_{\mathcal{S}_{s-p}}(\varphi_n(\rho^A_i\otimes\rho^B_i))^Rd\mu(\rho^A\otimes\rho^B)\\
&=\int_{\mathcal{S}_{s-p}}(\varphi(\rho^A_i\otimes\rho^B_i))^Rd\mu(\rho^A\otimes\rho^B)\end{array}\eqno{(29)}$$ with respect to the Hilbert-Schmidt norm.
Therefore, $$\begin{array}{rl}\|\rho^R\|_{\rm Tr}&\leq\int_{\mathcal{S}_{s-p}}\|(\varphi(\rho^A\otimes\rho^B))^R\|_{\rm Tr}d\mu(\rho^A\otimes\rho^B)\\
&=\int_{\mathcal{S}_{s-p}}1d\mu(\rho^A\otimes\rho^B)=1\end{array}$$
as $\|(\varphi(\rho^A\otimes\rho^B))^R\|_{\rm Tr}=1$ by Eq.(27).
This completes the proof. \hfill$\Box$

Let $\mathcal{S}_{sep}^R=\{\rho^R:\rho\in\mathcal{S}_{sep}\}$. The
previous criterion shows that
$\mathcal{S}_{sep}^R\subset\mathcal{T}_1(H_B\otimes H_B,H_A\otimes
H_A)$, where $\mathcal{T}_1(H_B\otimes H_B,H_A\otimes H_A)$ denotes
the set of all trace-class operators from $H_B\otimes H_B$ into
$H_A\otimes H_A$  with the trace-norm not greater than 1.

Next we will show that, there is another alternative way to perform
realignment operation for all states which is equivalent to the
operation proposed as in Definition 3.1.

Let $A\in\mathcal{B}(H_A)$, $B\in\mathcal{B}(H_B)$. For given bases
$\{|m\rangle\}$ and $\{|\mu\rangle\}$ of $H_A$ and $H_B$,
respectively, $A$ and $B$ can be written in the form
$A=\sum\limits_{m,n}a_{mn}|m\rangle\langle n|$ and
$B=\sum\limits_{\mu,\nu}b_{\mu\nu}|\mu\rangle\langle \nu|$. Regard
$A$ as a vector
$|A\rangle=\sum\limits_{m,n}a_{mn}|m\rangle|n\rangle$ in the Hilbert
space ${\mathcal C}_2(H_A)$
 and
$B$ as
$|B\rangle=\sum\limits_{\mu,\nu}b_{\mu\nu}|\mu\rangle|\nu\rangle$ in
the Hilbert space ${\mathcal C}_2(H_B)$, respectively. Let $\langle
B|$ denote the transpose of $|B\rangle$. Let $\rho$ be a
separable state with $\rho=\sum\limits_ip_i\rho_i^A\otimes\rho_i^B$,
where $p_i\geq0$, $\sum\limits_ip_i=1$, $\rho_i^A$ and $\rho_i^B$
are pure states in $\mathcal{S}(H_A)$ and $\mathcal{S}(H_B)$,
respectively. It is easy to see that
$\rho^R=\sum\limits_ip_iD_i\otimes\bar{D_i}
=\sum\limits_ip_i|\rho_i^A\rangle\langle\rho_i^B|$ as
$|\rho_i^A\rangle\langle\rho_i^B|=D_i\otimes\bar{D_i}$. This
motivates the possibility of generalizing Eq.(7) to
infinite-dimensional cases.

To do this, notice that $\mathcal{S}(H_A\otimes H_B)\subset\mathcal{T}(H_A\otimes
H_B)\subset\mathcal{C}_2(H_A\otimes
H_B)=\mathcal{C}_2(H_A)\otimes\mathcal{C}_2(H_B)$. So each state
$\rho$ can be regarded as a ``vector'' of the Hilbert space
$\mathcal{C}_2(H_A)\otimes\mathcal{C}_2(H_B)$. Considering the
Fourier representation of
$T\in\mathcal{C}_2(H_A)\otimes\mathcal{C}_2(H_B)$ with respect to a
product basis of $\mathcal{C}_2(H_A)\otimes\mathcal{C}_2(H_B)$, we
see that $T$ can be written in the form
$$T=\sum\limits_kA_k\otimes B_k,\eqno{(30)}$$ where
$\{A_k\}\subset\mathcal{C}_2(H_A)$,
$\{B_k\}\subset\mathcal{C}_2(H_B)$ and the series converges in the
Hilbert-Schmidt norm.

{\bf Proposition 3.4.} \ {\it Let $T\in\mathcal{C}_2(H_A\otimes
H_B)$. Write $T=\sum\limits_kA_k\otimes B_k$ as in Eq.(30). Then,
with respect to given bases $\{|m\rangle\}$ and $\{|\mu\rangle\}$ of
$H_A$ and $H_B$, respectively, we have
$$T^R=\sum\limits_k|A_k\rangle\langle B_k|,\eqno{(31)}$$
where the series converges in Hilbert-Schmidt norm on
$\mathcal{C}_2(H_B\otimes H_B,H_A\otimes H_A)$,
$|A_k\rangle=\sum\limits_{m,n}a_{mn}^{(k)}|m\rangle|n\rangle$ if
$A_k=\sum\limits_{m,n}a_{mn}^{(k)}|m\rangle\langle n|$,
$|B_k\rangle=\sum\limits_{\mu,\nu}b_{\mu\nu}^{(k)}|\mu\rangle|\nu\rangle$
if $B_k=\sum\limits_{\mu,\nu}b_{\mu\nu}^{(k)}|\mu\rangle\langle
\nu|$ and $\langle B_k|$ denotes the transpose of $|B_k\rangle$,
$k=1$, 2, $\dots$. }

{\bf Proof.} \  Write
$\mathcal{R}^\prime(T)=\sum\limits_k|A_k\rangle\langle B_k|$
whenever $T=\sum\limits_kA_k\otimes B_k$ as in Eq.(30). We show that
$$\mathcal{R}^\prime(T)=\mathcal{R}(T)=T^R\eqno{(32)}$$
holds for all $T\in\mathcal{C}_2(H_A\otimes H_B)$. It is easy to
check that, if $T=\sum\limits_k A_k\otimes B_k=(t_{m\mu,n\nu})$,
then $T^{R^\prime}=\sum\limits_k|A_k\rangle\langle
B_k|=(\tilde{t}_{mn,\mu\nu})$ with
$\tilde{t}_{mn,\mu\nu}=t_{m\mu,n\nu}$. Thus, $T^{R^\prime}=T^R$ is
well defined. It remains to show that the series
$\sum\limits_k|A_k\rangle\langle B_k|$ converges to $T$ in the
Hilbert-Schmidt norm. Let $T_n=\sum\limits_{k=1}^nA_k\otimes B_k$
and $T_n^R=\sum\limits_{k=1}^n|A_k\rangle\langle B_k|$; then
$\|T^R-T^R_n\|_2=\|\sum\limits_{k=n+1}^{\infty}|A_k\rangle\langle
B_k|\|_2=\|\sum\limits_{k=n+1}^{\infty}A_k\otimes
B_k\|_2\rightarrow0$ ($n\rightarrow+\infty$) since
$\|T-T_n\|_2\rightarrow0$ ($n\rightarrow+\infty$).\hfill$\Box$

By now, for any state $\rho\in\mathcal{S}(H_A\otimes H_B)$, we have three equivalent
definitions of the realignment operator of it.

Inspired by Eq.(30) and an idea in \cite{RO2}, we generalize the
notion of ``\emph{computable cross norm}'' to the
infinite-dimensional case.

{\bf Definition 3.5.} \ {\it The computable cross norm $\|T\|_{\rm
CCN}$ of an arbitrary element $T\in\mathcal{C}_2(H_A\otimes H_B)$ is
defined by
$$\|T\|_{\rm CCN}:=\inf\{\sum\limits_k\|
A_k\|_2\|B_k\|_2: T=\sum\limits_kA_k\otimes B_k\},\eqno{(33)}$$
where the infimum runs over all decompositions of $T$ into
elementary tensors as that in Eq.(30).}

It is evident that $\|\cdot\|_{\rm CCN}$ is a cross norm on
$\mathcal{C}_2(H_A\otimes H_B)$ since $\|A\otimes B\|_{\rm
CCN}=\|A\|_2\|B\|_2$ for all $A\in\mathcal{C}_2(H_A),
B\in\mathcal{C}_2(H_B)$. Also, we may have $\|T\|_{\rm CCN}=+\infty$
for some $T$.

 Noticing that,
every vector in the tensor product Hilbert space of two Hilbert
spaces has a so-called Schmidt decomposition \cite{GY}. Together
with the fact $\mathcal{C}_2(H_A\otimes
H_B)=\mathcal{C}_2(H_A)\otimes\mathcal{C}_2(H_B)$, we can derive
that, for any state $\rho$ on $H_A\otimes H_B$, $\rho$ has a Schmidt
decomposition  as a vector in $\mathcal{C}_2(H_A)\otimes\mathcal{C}_2(H_B)$, i.e.,
$$\rho=\sum\limits_{k=1}^{N_\rho}\delta_kE_k\otimes F_k,\eqno{(34)}$$ where
$E_k\in\mathcal{C}_2(H_A)$, $F_k\in\mathcal{C}_2(H_B)$ satisfying
${\rm Tr}(E_k^\dagger E_l)=\delta_{kl}$ and ${\rm Tr}(F_k^\dagger
F_l)=\delta_{kl}$, $k$, $l=1$, 2, $\dots, N_\rho$, the positive
scalars $\delta_1\geq \delta_2\geq \cdots $ are uniquely determined
by the corresponding vector $\rho$, and they are the so-called
Schmidt coefficients of $\rho$ \cite{GY}, while $N_\rho$ (may be
$+\infty$) is called the Schmidt number of $\rho$. Since ${\rm
Tr}(\rho^2)=\langle\rho|\rho\rangle=\sum\limits_k\delta_k^2$, we
have $\sum\limits_k\delta_k^2=1\Leftrightarrow\rho$ is a pure state
and $\sum\limits_k\delta_k^2<1\Leftrightarrow\rho$ is a mixed state.

The following lemma  highlights the relations among the trace norm
of the realignment operator, the computable cross norm and the sum
of the Schmidt coefficients of a state. As one might expect, the result
is the same as that for the finite-dimensional case.

{\bf Lemma 3.6.} \ {\it Let $\rho$ be a state in
$\mathcal{S}(H_A\otimes H_B)$ and $\{\delta_k\}$ be the Schmidt
coefficients of $\rho$  as a vector in $\mathcal{C}_2(H_A)\otimes\mathcal{C}_2(H_B)$. Then we have
$$\|\rho^{R}\|_{\rm Tr}=\|\rho\|_{\rm
CCN}=\sum\limits_k\delta_k.\eqno{(35)}$$}

{\bf Proof.} \ Let $\rho=\sum\limits_kA_k\otimes B_k$ as in Eq.(30)
and $\rho=\sum\limits_k\delta_kE_k\otimes F_k$ be the Schmidt
decomposition of $\rho$ as a vector in $\mathcal{C}_2(H_A)\otimes\mathcal{C}_2(H_B)$, where the series converges in
Hilbert-Schmidt norm. Then $\rho^{R}=\sum\limits_k|A_k\rangle\langle
B_k|$, and $$\|\rho^{ R}\|_{\rm
Tr}=\sum\limits_k\delta_k\eqno{(36)}$$ since one can regard
$\rho^{R}=\sum\limits_k\delta_k|E_k\rangle\langle F_k|$ as the
singular value decomposition of $\rho^{R}$. Next, we  show that
$$\|\rho^R\|_{\rm Tr}
=\inf\{\sum\limits_k\||A_k\rangle\|\cdot\||B_k\rangle\|:\rho
=\sum\limits_kA_k\otimes B_k\}.\eqno{(37)}$$ On the one hand, we
have $\|\rho^R\|_{\rm Tr}\leq\sum\||A_k\rangle\langle B_k|\|_{\rm
Tr}=\sum\limits_k\||A_k\rangle\|\cdot\||B_k\rangle\|$. On the other
hand, $\|\rho^R\|_{\rm
Tr}=\sum\limits_k\delta_k\||E_k\rangle\|\cdot\||F_k\rangle\|=\sum\limits_k\delta_k$
since $\||E_k\rangle\|=\|E_k\|_2=\||F_k\rangle\|=\|F_k\|_2=1$.
Namely, the infimum is attained at the singular value decomposition
of $\rho$. Now, we arrive at $$\begin{array}{rl}\|\rho^{R}\|_{\rm
Tr}=&\inf\{\sum\limits_{k}\||A_k\rangle\|\cdot\||B_k\rangle\|
:\rho=\sum\limits_kA_k\otimes
B_k\}\\
=&\inf\{\sum\limits_{k}\|A_k\|_2\|B_k\|_2:\rho=\sum\limits_kA_k\otimes
B_k\}\\=&\|\rho\|_{\rm CCN},\end{array}\eqno{(38)}$$ which completes
the proof.\hfill$\Box$

For a pure state $\rho_\psi\in\mathcal{S}(H_A\otimes H_B)$, let
$|\psi\rangle=\sum\limits_k\lambda_k|m_k\rangle|\mu_k\rangle$ be the
Schmidt decomposition of $|\psi\rangle\in H\otimes K$, then it is straightforward that
$$\|\rho_\psi\|_{\rm CCN}=(\sum_k\lambda_k)^2.\eqno{(39)}$$
From this it is obvious that
 a pure state $\rho$
is separable if and only if $\|\rho^R\|_{\rm Tr}=\|\rho_\psi\|_{\rm CCN}=1$. Further more, combining
Criterion 3.3 and Lemma 3.6,
 we establish the RCCN criterion for
the infinite-dimensional systems,  that is, the criterion below
is the main result of this paper.

{\bf Criterion 3.7.} ({\bf The RCCN criterion for
infinite-dimensional bipartite quantum systems}) \ {\it Let
$\rho\in\mathcal{S}(H_A\otimes H_B)$ and $\{\delta_k\}$ be  the
Schmidt coefficients of $\rho$  as a vector in $\mathcal{C}_2(H_A)\otimes\mathcal{C}_2(H_B)$. If $\rho$ is separable, then
$$\|\rho^{R}\|_{\rm Tr}=\|\rho\|_{\rm
CCN}=\sum\limits_k\delta_k\leq1.\eqno{(40)}$$ In particular, assume
that $\rho$ is a pure state, then $\rho$ is separable if and only if
$$\|\rho^{R}\|_{\rm
Tr}=\|\rho\|_{\rm CCN}=\sum\limits_k\delta_k=1.\eqno{(41)}$$}

In what follows, we give some  examples to illustrate that  there
exist PPT entangled states that can be detected by the RCCN
criterion, there exist PPT entangled states that can not be detected
by the realignment criterion and there exist non-PPT entangled state
that can not be detected by the RCCN criterion. These examples imply
that the RCCN criterion is neither `weaker' nor `stronger' than the
PPT criterion. There also exist entangled states which can not be
detected by any one of these two criteria. However, we can show
that, for the so-called `symmetric sates', the PPT criterion is
equivalent to the RCCN criterion, namely, a symmetric state $\rho$
satisfying $\|\rho^R\|_{\rm Tr}\leq1$ if and only if it is a PPT
state (see in Proposition 3.13).

{\bf Example 3.8.} \ Let $H_A$ and $H_B$ be complex Hilbert spaces
with orthonormal bases $\{|0\rangle$, $|1\rangle$, $\dots\}$ and
$\{|0^\prime\rangle$, $|1^\prime\rangle$, $\dots\}$, respectively.
Let $$\rho_{\alpha} =\frac{2}{7}|w\rangle\langle
w|+\frac{\alpha}{7}\sigma_+ +\frac{5-\alpha}{7}\sigma_-,$$ where
$|w\rangle=\frac{1}{\sqrt{3}}(|0\rangle|0^\prime\rangle+|1\rangle|1^\prime\rangle+|2\rangle|2^\prime\rangle)$,
$\sigma_+=\frac{1}{3}(|0\rangle|1^\prime\rangle\langle 0|\langle
1^\prime|+ |1\rangle|2^\prime\rangle\langle 1|\langle2^\prime|
+|2\rangle|0^\prime\rangle\langle 2|\langle0^\prime|)$,
$\sigma_-=\frac{1}{3}(|1\rangle|0^\prime\rangle\langle
1|\langle0^\prime|+|2\rangle|1^\prime\rangle\langle
2|\langle1^\prime| +|0\rangle|2^\prime\rangle\langle
0|\langle2^\prime|)$ and $2\leq\alpha\leq5$. A straightforward
calculation shows that $$\|\rho_\alpha^R\|_{\rm
Tr}=\frac{19}{21}+\frac{2}{21}\sqrt{19-15\alpha+3\alpha^2}.$$ It is
easy to check that $\|\rho_\alpha^R\|_{\rm Tr}\leq1$ if and only if
$2\leq\alpha\leq3$. Thus, by the RCCN criterion, $3<\alpha\leq5$
implies $\rho_\alpha$ is entangled.

Define $$\sigma=\sum\limits_{i=3}^{+\infty}p_i|i\rangle\langle
i|\otimes|i^\prime\rangle\langle i^\prime|,\ p_i\geq0,\
\sum\limits_{i=3}^{\infty}p_i=1.$$ Then $\sigma$ is a separable
state and $\|\sigma^R\|_{\rm Tr}=1$. Let
$$\rho_{t,\alpha}=t\rho_\alpha+(1-t)\sigma, \ 0<t\leq1, 3<
\alpha\leq 4.\eqno{(42)}$$ By \cite{RO2,QH}, it is easily checked that
$\rho_{t,\alpha}$ is a PPT state whenever $3<\alpha\leq 4$ since $\rho_{\alpha}^{T_B}\geq0$ whenever $3<\alpha\leq4$. On
the other hand,
$$\|\rho_{t,\alpha}^R\|_{\rm Tr}=\|t\rho_\alpha^R+(1-t)\sigma^R\|_{\rm Tr}=
t\|\rho_\alpha^R\|_{\rm Tr}+(1-t)\|\sigma^R\|_{\rm Tr}.$$ It follows
that $\|\rho_{t,\alpha}^R\|_{\rm Tr}>1$ for all $0\leq t<1$ as
$\|\sigma^R\|_{\rm Tr}=1$ and $\|\rho_\alpha^R\|>1$ whenever
$3<\alpha\leq 4$. So, there are PPT entangled states that can be
detected by the RCCN criterion.

The example below is discussed in \cite{HJ2} and it illustrates
particularly that there exist non-PPT states as well as entangled
PPT states that con not be recognized by the RCCN criterion.

{\bf Example 3.9.}  Let $H_A$ and $H_B$ be complex Hilbert spaces of
dimension $\geq 4$ with orthonormal bases $\{|0\rangle,|1\rangle$,
$|2\rangle$, $\dots\}$ and $\{|0'\rangle, |1^\prime\rangle$,
$|2^\prime\rangle$, $\dots\}$, respectively. Let
$|\omega\rangle=\frac{1}{\sqrt{4}}(|0\rangle|0'\rangle+|1\rangle|1'\rangle+|2\rangle|2'\rangle
+|3\rangle|3'\rangle).$ Define
$\rho_1=|\omega\rangle\langle\omega|$,
$\rho_2=\frac{1}{4}(|0\rangle|1'\rangle\langle0|\langle
1'|+|1\rangle|2'\rangle\langle1|\langle
2'|+|2\rangle|3'\rangle\langle2|\langle 3'|
+|3\rangle|0'\rangle\langle3|\langle0'|),$
$\rho_3=\frac{1}{4}(|0\rangle|2'\rangle\langle0|\langle2'|
+|1\rangle|3'\rangle\langle1|\langle3'|+|2\rangle|0'\rangle\langle2|\langle0'|+|3\rangle|1'\rangle\langle3|\langle1'|)$
and $\rho_4=\frac{1}{4}(|0\rangle|3'\rangle\langle0|\langle3'|
+|1\rangle|0'\rangle\langle1|\langle0'|+|2\rangle|1'\rangle\langle2|\langle1'|+|3\rangle|2'\rangle\langle3|\langle2'|).$
Let $$\rho=\sum_{i=1}^4q_i\rho_i \quad \mbox{and}\quad
\rho_t=(1-t)\rho+t\rho_0,\eqno(43)$$ where
 $q_i\geq 0$ for $i=1,2,3,4$ with $q_1+q_2+q_3+q_4=1$, $t\in[0,1]$,
and $\rho_0$ is a state on $H_A\otimes H_B$. It was shown in
\cite{QH} that, for sufficiently small $t$, or for $\rho_0$ with
$|i\rangle|\mu'\rangle\langle
j|\langle\nu'|\rho_0=\rho_0|i\rangle|\mu'\rangle\langle
j|\langle\nu'|=0$ for any $i,j,\mu,\nu\in\{0,1,2,3\}$, the following
statements are true.

(1) If $q_i<q_1$ for some $i=2,3,4$, then $\rho_t$ is entangled.

(2)   Let $\rho_0$ be PPT. Then $\rho_t$ is PPT if and only if
$q_2q_4\geq q_1^2$ and $q_3\geq q_1$. Thus,  if
$0<q_i<q_1<\frac{1}{4}$, $\frac{1}{4}\leq q_j<1$ with $q_iq_j\geq
q_1^2$ and $0<q_1\leq q_3<1$, where $i,j\in\{2,4\}$ and $i\not=j$,
then $\rho_t$ is PPT entangled.

(3) The trace norm of the realignment operator of $\rho$ is
$$\begin{array}{rl}\|\rho^R\|_{\rm Tr}=&\frac{3}{4}\sqrt{\sum_{i=1}^4q_i^2-q_1q_2-q_2q_3-q_3q_4-q_1q_4}\\
&+\frac{1}{4}\sqrt{\sum_{i=1}^4q_i^2+3(q_1q_2+q_2q_3+q_3q_4+q_1q_4)}+3q_1.\end{array}$$
Thus, if $\rho_0$ is PPT, and if $q_1\geq\frac{1}{6}$,
$q_i=\frac{1}{2}q_1$, $q_j=\frac{1}{2}$ and $q_3=\frac{1}{2}-3q_i$,
where $i,j\in\{2,4\}$ and $i\not=j$, then $\rho_t$ is PPT entangled
with $\|\rho^R\|_{\rm Tr}>1$ for sufficient small $t$, that is,
$\rho_t$ is PPT entangled that can be detected by the RCCN
criterion; if $q_1\leq\frac{1}{7}$, $q_i=\frac{1}{2}q_1$,
$q_j=\frac{1}{2}$ and $q_3=\frac{1}{2}-3q_i$, where $i,j\in\{2,4\}$
and $i\not=j$, then, for sufficient small $t$, $\|\rho^R\|_{\rm
Tr}<1$ and $\rho_t$ is PPT entangled but can not be detected by the
RCCN criterion.

The following illustrates that how to find suitable $\rho_t$ so that
$\|\rho^R_t\|_{\rm Tr}<1$ but $\rho_t$ is not PPT.

If $\rho_0$ is not PPT, we choose $q_1\leq\frac{1}{7}$,
$q_i=\frac{1}{2}q_1$, $q_j=\frac{1}{2}$ and $q_3=\frac{1}{2}-3q_i$,
where $i,j\in\{2,4\}$ and $i\not=j$. Then, as mentioned above we
have $\|\rho^R\|_{\rm Tr}<1$. Thus,  for sufficient small $t$,
$\|\rho^R_t\|_{\rm Tr}<1$. This means that $\rho_t$ is not PPT but
can not be recognized by the RCCN criterion.

If $\rho_0$ is PPT, we choose $q_2=\frac{1}{2}-\frac{3}{2}q_1$,
$q_3=\frac{1}{2}q_1$ and $q_4=\frac{1}{2}$. A computation shows that
$\|\rho^R\|_{\rm Tr}<1$ for $q_1\leq\frac{1}{7}$. For instance,
$\|\rho^R\|_{\rm Tr}=0.9866$ if $q_1=\frac{1}{7}$; $\|\rho^R\|_{\rm
Tr}=0.9496$ if $q_1=\frac{1}{8}$; $\|\rho^R\|_{\rm Tr}=0.7264$ if
$q_1=\frac{1}{100}$. Since $q_3<q_1$, $\rho$ is not PPT and thus
$\rho_t$ is not PPT. However, $\|\rho^R_t\|_{\rm Tr}<1$ for
sufficient small $t$.

{\bf Example 3.10.}\ Let $H_A=H_B$ be complex Hilbert spaces with
orthonormal bases $\{|0\rangle$, $|1\rangle$, $\dots\}$. Fixing a
positive number $3\leq m\in\mathbb{N}$. Define
$$\rho_{m,c}:=\frac{1}{m^3-m}((m-c)P_m+(mc-1)F_m),$$ where
$P_m:=\sum\limits_{i=0}^{m-1}\sum\limits_{j=0}^{m-1}|i\rangle\langle
i|\otimes|j\rangle\langle j|$ and
$F_m:=\sum\limits_{i=0}^{m-1}\sum\limits_{j=0}^{m-1}|i\rangle\langle
j|\otimes|j\rangle\langle i|$. It is easy to check  that
$$\|\rho_{m,c}^R\|_{\rm Tr}=\left\{\begin{array}{rl}\frac{2}{m}-c:&\ \ {\rm if}\ \ -1\leq c\leq\frac{1}{m},\\
c:&\ \ {\rm if}\ \ 1\geq c\geq\frac{1}{m}.\end{array}\right.$$ If
$\dim H_A=m$, then $\rho_{m,c}=\rho_c$ is the so-called Werner state
\cite{HP2}. It is shown in \cite{HP2} that $$\rho_{c}\ \mbox{\rm is
separable}\Leftrightarrow\rho^{T_B}\geq0\Leftrightarrow 0\leq c\leq
1.$$ We can derive that $$\rho_{m,c}\ \mbox{\rm is
separable}\Leftrightarrow\rho^{T_B}\geq0\Leftrightarrow0\leq c\leq
1.$$ Consequently, if $\frac{2}{m}-1\leq c<0$, then $\rho_{m,c}$ is
a non-PPT entangled state which satisfies $\|(\rho_{m,c})^R\|\leq
1$, that is, $\rho_{m,c}$ can not be recognized by the RCCN
criterion.

Let $$\varrho=\sum\limits_{i=m}^{+\infty}p_i|i\rangle\langle
i|\otimes|i\rangle\langle i|,\ p_i\geq0,\
\sum\limits_{i=m}^{+\infty}p_i=1,$$ then $\varrho$ is separable. Let
$$\rho_{\varepsilon,c}=\varepsilon\varrho+(1-\varepsilon)\rho_{m,c},\
0\leq\varepsilon<1, \ \frac{2}{m}-1\leq c<0.\eqno{(44)}$$ It is
straightforward that $\rho_{\varepsilon,c}$ is a state acting on
$H_A\otimes H_B$ and $\rho_{\varepsilon,c}^{T_B}$ is not positive. Now,
we can conclude that for any $\frac{2}{m}-1\leq c<0$ and
$0\leq\varepsilon<1$, $\rho_{\varepsilon,c}$ is a non-PPT
entangled state satisfying $\|\rho_{\varepsilon,c}^R\|_{\rm Tr}\leq1$ since $\|\varrho^R\|_{\rm Tr}=1$ and $\|\rho_{m,c}^R\|_{\rm Tr}\leq1$.

Now let us turn to another related topic. In \cite{TG}, several
entanglement criteria for the so-called `symmetric states' in the
finite-dimensional bipartite quantum systems are presented. Recall
that, a state $\rho$ on a finite dimensional bipartite system
$H_A\otimes H_B$ is called a symmetric state if $\dim H_A=\dim
H_B=N$ and $\rho=F\rho=\rho F$, where $F$ is the flip operator,
namely, $F$ satisfies
$F|\psi_A\rangle|\psi_B\rangle=|\psi_B\rangle|\psi_A\rangle$ for any
$|\psi_A\rangle\in H_A$ and $|\psi_B\rangle\in H_B$. It is showed
that, for a symmetric state $\rho$ in a finite-dimensional bipartite
quantum system, $\|\rho^R\|_{\rm Tr}\leq1$ if and only if $\rho$ is
a PPT state. Inspired by \cite{TG}, we can generalize the conception
of the symmetric states to the infinite-dimensional case with the
same spirit.

{\bf Definition 3.11.}\ {\it Let $H_A$ and $H_B$ be Hilbert spaces
with $\dim H_A=\dim H_B=+\infty$. Let $\{|m\rangle\}$ and
$\{|\mu\rangle\}$ be orthonormal bases respectively of $H_A$ and
$H_B$. A state $\rho\in\mathcal{S}(H_A\otimes H_B)$ is said to be
symmetric if
$$\rho=F\rho=\rho F,\eqno{(45)}$$ where
$F=\sum\limits_{m,\mu}|m\rangle|\mu\rangle\langle\mu|\langle m|$.}

The operator $F$ in the definition is called the flip operator. It
is clear that
$F|\psi_A\rangle|\psi_B\rangle=|\psi_B\rangle|\psi_A\rangle$ for any
$|\psi_A\rangle\in H_A$, $|\psi_B\rangle\in H_B$.

 Write
$\rho=(\rho_{m\mu,n\nu})$, where $\rho_{m\mu,n\nu}=\langle
m|\langle\mu|\rho|n\rangle|\nu\rangle$, one can obtain

{\bf Lemma 3.12.}\ {\it If $\rho\in\mathcal{S}(H_A\otimes
H_B)$ is a symmetric state,  then
$$\rho_{m\mu,n\nu}=\rho_{\mu m,n\nu}=\rho_{m\mu,\nu n}=\rho_{\mu
m,\nu n}.\eqno{(46)}$$ Moreover, $$F\rho^R=\rho^{T_A}.\eqno{(47)}$$}

{\bf Proof.} Write $\rho^R=(\hat{\rho}_{m\mu,n\nu})$,
$F\rho^R=(\check{\rho}_{m\mu,n\nu})$. It turns out that
$\hat{\rho}_{m\mu,n\nu}=\rho_{mn,\mu\nu}$,
$\check{\rho}_{m\mu,n\nu}=\hat{\rho}_{\mu m,n\nu}$, and thus
$\check{\rho}_{m\mu,n\nu}=\hat{\rho}_{\mu m,n\nu}=\rho_{\mu
n,m\nu}$.   On the other hand, writing
$\rho^{T_A}=(\tilde{\rho}_{m\mu,n\nu})$, we have
$\tilde{\rho}_{m\mu,n\nu}=\rho_{n\mu,m\nu}$. Therefore,
$\rho^{T_A}=F\rho^R$ since $\rho_{\mu
n,m\nu}=\rho_{n\mu,m\nu}$.\hfill$\Box$

As $F$ is  unitary, by Lemma 3.12, the singular values of $\rho^R$
is equal to the singular values of $\rho^{T_A}$. Since ${\rm
Tr}(\rho^{T_A})=1$, it follows that $\rho^{T_A}$ is not positive if
and only if $\rho^{T_A}$ has at least one negative eigenvalue.
Therefore, $\rho$ is not PPT implies that $\|\rho^R\|_{\rm
Tr}=\|\rho^{T_A}\|_{\rm Tr}>1$ and vice versa. Thus we have proved
the following

{\bf Proposition 3.13.}\ {\it If $\rho\in\mathcal{S}(H_A\otimes
H_B)$ is symmetric, then $\rho$ is a PPT state if and only if
$\|\rho^R\|_{\rm Tr}\leq1$.}

\section{Conclusion}

In conclusion, we generalize the row realignment operation, the
Computable Cross Norm to the states   of   infinite-dimensional
bipartite quantum systems. The realignment operators of the states
are Hilbert-Schmidt operators from $H_B\otimes H_B$ into $H_A\otimes
H_A$, and moreover, the row realignment operation $T\mapsto T^R$ is
an isometric linear map from $\mathcal{C}_2(H_A\otimes H_B)$ into
$\mathcal{C}_2(H_B\otimes H_B, H_A\otimes H_A)$. Similar to that in
the finite-dimensional bipartite quantum systems, there are two
kinds of realignment operations, namely, the row realignment
operation and the column realignment operation. These two
realignment operations are equivalent up to the trace norm. So, it
suffices to discuss the row realignment operation.

In fact, three equivalent definitions of the realignment operation
are introduced. This allow us to establish the realignment criterion
and the RCCN criterion of separability for states in
infinite-dimensional bipartite systems. Thus, for both
finite-dimensional and infinite-dimensional systems, if a state
$\rho\in{\mathcal S}(H_A\otimes H_B)$ is
 separable, then $\|\rho^R\|_{\rm Tr}=\|\rho\|_{\rm
 CCN}=\sum_k\delta_k\leq 1$, where $\|\rho\|_{\rm
 CCN}$ is the computable cross norm of $\rho$ and $\{\delta_k\}$ are the Schmidt coefficients of $\rho$  as a vector
 in the Hilbert space ${\mathcal C}_2(H_A)\otimes {\mathcal C}_2(H_B)$. For the case that $\rho$ is pure,
 $\rho$ is   separable if and only if $\|\rho^R\|_{\rm Tr}=\|\rho\|_{\rm
 CCN}=\sum_k\delta_k= 1$.
Like the case of finite-dimension,  the RCCN criterion and the PPT
criterion are independent as illustrated by examples, and   these
two criteria are equivalent for the symmetric states.

{\bf Acknowledgement.} This work is partially supported by the
National Natural Science Foundation of China (10771157)  and the
Research Fund of Shanxi for Returned Scholars (2007-38).

\end{document}